# Evidence of large anisotropy in the magnetization of $Na_{0.35}CoO_2.1.3H_2O$ quasi-single-crystal superconductors

P. Badica, [1,2] T. Kondo, [1] K. Togano [1] and K. Yamada [1]

[1]Institute for Materials Research, Tohoku University, 2-1-1 Katahira,

Aoba-ku, Sendai, 980-8577, Japan

[2]National Institute for Materials Physics, P.O. Box MG-7, 76900 Bucharest, Romania

Quasi-single crystals (up to 2x2x1 $mm^3$) of $Na_{0.35}CoO_2.1.3H_2O$-superconductor have been grown. Magnetization M(H, T) and M(T, H) curves with magnetic field approximately parallel and perpendicular to c-axis indicates on large anisotropy, comparable with Bi-based high-temperature superconducting (HTS) phases.

PACS numbers: 74.25.Ha  74.25.Dw  74.25.Fy  74.70.-b

Recently, Takada et al. [1] reported superconductivity in $Na_{0.35}CoO_2.1.3H_2O$ layered oxide material without Cu. Material resembles HTS from several points of view and at the same time some differences have been noted [1-3]. Understanding of these features vs. HTS is of great interest in determination of the design principles of new superconductors, in improvement of the second ones targeting different practical applications or generally in better understanding of superconductivity.

Some of the superconducting phenomenological parameters were deduced in previous works [2] on randomly aligned powdered samples. Large anisotropy in superconducting properties has been anticipated and assignment of this material to unconventional extreme type II superconductors has been suggested [4]. Further step would be analysis of magnetic and/or electrical properties on different directions of the aligned samples or single crystals for the determination of the superconducting anisotropy. Lack of single crystals and low stability of material in normal atmospheric conditions [5] or in different media such as resins or gels usually used to retain alignment under high magnetic fields is making this task extremely difficult.

In this letter we report growth of quasi-single crystals and magnetization M(H, T) and M(T, H) data emphasizing superconducting anisotropy of the material. SEM images will be also presented.

Single crystal $Na_{0.71}Co_{0.96}O_2$ has been grown by traveling melting zone method using a mirror type furnace. A rod of 5mm in diameter and 50mm in length was obtained. A cross section along the length of the rod is showing existence of the hexagonal domains in good agreement with XRD (Rigaku Rint 2000, $Cu_{K\alpha}$ radiation) that indicates a hexagonal crystal lattice (JCPDS 30-1182). It results that the c-axis lays in the circular cross section of the crystal rod and the growth direction is *a*-axis direction (along the length of the rod). Next, a disc of approximately 1mm thickness has been cut along *c*-axis and immersed in $Br_2$ with a concentration of 6M in acetonitrile, for 28.5hrs, at room temperature. Sample has been washed in water and acetonitrile and was kept for $H_2O$ insertion for 135 hrs using the same arrangement as for powder samples [6]. Final product consists of several almost equiaxial pieces of approximately 1cm$^3$ or more and powder.

SEM images (JEOL JSM-6700) from Fig. 1a, b shows the layered nature of the superconducting crystals. Parallel *ab*-layers are separated by cracks. Cracks are the path through which Na-extraction and $H_2O$-insertion take place. The morphology of the crystals is not perfect. This issue is of key importance in determination of the anisotropy of the material. Generally, misalignment of the *ab*-layers from one quasi crystal is up to 10° and layers have wavy shape. On the top or bottom terminal *ab*-layers, hexagonal domains, preserved from the parent material, can be easily visualized (Fig. 1c). Based on this observation, each piece can be considered an aligned structure or a quasi-single crystal. XRD data are supporting this result; XRD pattern taken on the flat surface of a crystal is showing diffraction lines ascribed to (00l) planes [1] (lattice parameter $c$=19.613 Å). Also (00l) peaks of the un-reacted parent material were detected. We believe the residual parent phase is necessary to keep the integrity of the aligned superconducting layered structure.

We have attempted to measure M(H, T) loops and M(T, H) curves in zero-field arrangement (ZFC) with applied magnetic field H parallel to *c*-axis and *ab*-plane (MPMS-XL5 SQUID magnetometer, Quantum Design). Although we shall use notation H//c and H//ab, alignment of the quasi crystals on the indicated directions is not perfect due to the observed defects. In fact, we have aligned our samples relative to the surface of the crystals. In this regard, crystals have been carefully selected under an optical microscope. During this procedure samples were kept on a Cu-plate placed on an ice block. Same method was applied when changing the alignment of the sample. After repositioning the sample in the magnetometer, ZFC/FC curves were measured at 20Oe to determine weather any modifications occurred in the behavior of the sample. No changes were observed. At the end of measurements on one direction, ZFC/FC curves were again measured and again no changes were detected. We conclude that the material is stable at low temperatures.

Magnetic measurements were performed on a single sample composed of 3 quasi-crystals with a total weight of 9.6mg. Magnetization loops and M(T, H) curves are presented in Figs. 2 and 3.

A general observation is that anisotropy is extremely large. This statement is based on the distinctive features of the M(H) curves (Fig. 2). Signal in magnetization for H//c around maximum M, located at low magnetic fields, is more than one order of magnitude larger than that for the other direction with H//ab. The difference in the magnitude of the signal for different directions of H is detected in ZFC-M(T) curves, either (Fig.3). This suggests that critical current $J_c$ flowing along c-axis direction, i.e. across *ab*-plane, is much smaller than $J_c$ in *ab*-plane, due to the layered structure of the material, as indicated by Kishio et al [7] for La-Sr-Cu-O single crystals with perfect shape. Same idea seems to work in our situation. But, the observed cracks are introducing some degree of uncertainty in the presented scenario in the sense that demagnetization and/or other effects related to contacts between the layers cannot be totally excluded to explain the signal difference. More experiments on cracks formation and cracks influence vs. magnetic properties are needed. Another observation is that the shape of M(H) curves is completely different for H//ab and H//c. One of the most interesting differences is that hysteresis for H//c rapidly disappears, while it remains up to much higher magnetic fields for H//ab (see also next paragraph). It results that our crystals have strong flux pinning force when placed in magnetic field parallel to *ab*-plane and this is probably due to the intrinsic pinning proposed by Tachiki et al [8]. In summary, our data have a high degree of similarity with many results reported [e.g. 7, 9-11] for HTS single crystals, leading to the conclusion that $Na_{0.35}CoO_2 \cdot 1.3H_2O$ superconductor has an extremely high 2D dimensionality. The closest matching considering specific features of the M(H) curves seems to be with Bi-2212 HTS [10, 11].

From M(H) loops (Fig. 2) we have estimated lower critical field $H_{c1}^{H//c}$ and irreversibility fields $H_{irr}^{H//c}$ and $H_{irr}^{H//ab}$. $H_{c1}$ values were taken as the field for which M(H) curve starts to deviate from the linear behavior generated by perfect diamagnetism. $H_{c1}^{H//c}$ data are plotted vs. temperature in Fig. 4a. Fitting with $H_{c1}=H_{c1}(0)[1-(T/T_c)^2]$ is resulting in $H_{c1}^{H//c}(0)$=17 Oe and $T_c$=3.8K. Considering relatively low number of $H_{c1}$-T points, the value of $T_c$ from fitting agrees well with the experimental value of 3.7K. For H//ab, it was not possible to identify a linear region in M(H) curves even when measuring with the highest resolution of our equipment (0.1Oe/step for H and down to 1.9K). We believe $H_{c1}^{H//ab}$ is taking low values as for HTS with high anisotropy (e.g. Bi-based phases). Irreversibility fields determined for a criterion of ΔM=0.005emu/g vs. temperature are presented in Fig. 4b. ΔM is the difference between the magnetization $M^+$ on the field increasing and decreasing $M^-$ branches of M(H) hysteresis. At 1.9 and 2.5K the anisotropy in irreversibility field ($H_{irr}^{H//ab}$ / $H_{irr}^{H//c}$) is 8 and 25, respectively. As reference, $H_{irr}^{H//c}$-T data, for a criterion of ΔM→0, are also presented in Fig. 4b.

Upper critical fields $H_{c2}^{H//c}$ and $H_{c2}^{H//ab}$ were determined considering the onset point of the superconducting transition for M(T) curves (Fig. 3). Variation of $H_{c2}$ vs.

temperature is shown in Fig. 4c. Experimental points were fitted with linear functions taking at T=0K values of 22T and 5.5T when H//ab and H//c, respectively. Upper critical fields for each direction were calculated by applying Werthamer-Hefland-Hohemberg (WHH) formula [12] for the type II superconductors in the dirty limit: $H_{c2}(0)=0.691 \times (dH_{c2}/dT)T_c$, with $dH_{c2}^{H//c}/dT=1.41$ K/T and $dH_{c2}^{H//ab}/dT=6.12$ K/T being the absolute slopes of the linear fitting of $H_{c2}(T)$ from Fig. 4c. Coherence length $\xi$, penetration depth $\lambda$, Ginzburg-Landau (GL) parameter $\kappa$ and anisotropy coefficient $\gamma$ were calculated from the relations: $H_{c2}^{H//c}=\Phi_0/2\pi\xi_{ab}^2$, $H_{c2}^{H//ab}=\Phi_0/2\pi\xi_{ab}\xi_c$, $H_{c1}^{H//c}=(\Phi_0/4\pi\lambda_{ab}^2)\ln(\lambda_{ab}/\xi_{ab})$, $\kappa^{H//c}=\lambda_{ab}/\xi_{ab}$, $\kappa^{H//ab}=\gamma\kappa^{H//c}$ and $\gamma=H_{c2}(0)^{H//ab}/H_{c2}(0)^{H//c}$. No demagnetization factors were considered. Results are gathered in Table 1.

Experimental value of the superconducting anisotropy determined in this work is close to the values of $\gamma$ usually reported for HTS with low anisotropy, such as Y-123. Due to misalignment in the sample, this value is likely to be significantly under-evaluated. Literature data on ceramic superconductors with layered structure (e.g. HTS, $MgB_2$) is showing that $\gamma$ determined on aligned powdered samples, i.e. on samples with some degree of misalignment, is taking values of approximately 3-6 times lower than the values measured for almost perfect single crystals. In our situation, quasi crystals are more like aligned powders than perfect single crystals. Hence, real $\gamma$-value is expected to be 3-6 times higher than the experimental value, i.e. within 12-25, or around an average value of 18. Similar high values for $\gamma$ are often reported for the Bi-based HTS system from magnetization measurements.

Another important piece of information is that $H_{c2}^{H//ab}(0)$ is of about 15T. This value is approximately 4 times lower than the $H_{c2}$ value deduced for randomly aligned powder samples with maximum reported critical tempearature ($T_c$=4.6K) [2]. More and precise measurements at lower temperatures are necessary.

We would like to emphasize that effort to try to align powder samples of $Na_{0.35}CoO_2.1.3H_2O$ superconductor in high magnetic fields is meaningless, unless $T_c$ of the samples would be above the value of 3.7K from this article. This is because of misalignment: generally, aligned powders are leading to under evaluated anisotropy values and, in this particular case, particles of $Na_{0.35}CoO_2.1.3H_2O$ - powder have shown the same wavy layered morphology containing cracks [6] as present quasi-single crystals. New methods for synthesis of perfect single crystals are required.

We have successfully synthesized quasi-crystals and investigated them by magnetization measurements. Phenomenological parameters were deduced and superconducting anisotropy $\gamma$ was determined. Experimental value of this parameter, $\gamma$=4.3, is under evaluated and M(H) loops suggests that $Na_{0.35}CoO_2.1.3H_2O$ is more like Bi-based HTS with high anisotropy than the value itself is indicating.

Authors would like to thank E. Aoyagi and Y. Hayasaka of HVEM Laboratory, Tohoku University for help with SEM measurements.

Figure and Table Caption

Figure 1 SEM images of a) two $Na_{0.35}CoO_2.1.3H_2O$ quasi-single crystals with different alignment b) detail showing the parallel wavy layered structure separated by cracks and c) detail taken on flat terminal *ab*-plane showing hexagonal domains.

Figure 2 M(H) loops measured at 1.9, 2.1, 2.3 and 2.5 K with a) H//ab and b) H//c.

Figure 3 M(T) curves in ZFC arrangement for a) H//ab (H=0.002, 0.1, 0.65, 1, 1.5, 2, 2.5, 3.5 and 5 T) and b) H//c (H=0.002, 0.1, 0.65, 1 and H=1.5, 2 and 2.5 T in inset). Arrows are indicating the increasing applied magnetic field. Curves measured at 20 Oe can be directly compared in a) and b) since in this case the units on y-axis are in emu/g. The other curves were shifted on y-axis for easy visualization.

Figure 4 Critical fields and irreversibility fields vs. temperature: a) -$H_{c1}$, b) -$H_{irr}$ c) -$H_{c2}$. Circles in Fig. 4b are data points of $H_{irr}^{H//c}$ for a criterion with $\Delta M=(M^+-M^-)\rightarrow 0$.

Table 1 Superconductivity parameters for $Na_{0.35}CoO_2.1.3H_2O$ sample.

Table1

| Lower critical field, $H_{c1}(0)$ (Oe) | Upper critical field, $H_{c2}(0)$ (T) | Coherence length, $\xi$ (nm) | Anisotropy factor, $\gamma$ - | Penetration depth, $\lambda$ (nm) | GL parameter, $\kappa$ - |
|---|---|---|---|---|---|
| - | $H_{c2}^{H//ab}(0)= 15.6$ | $\xi_{ab}= 9.6$ | | $\lambda_{ab}= 641$ | $\kappa^{H//c}= 66.7$ |
| $H_{c1}^{H//c}(0)= 17$ | $H_{c2}^{H//c}(0)= 3.6$ | $\xi_c= 2.23$ | 4.3 | $\lambda_c=\gamma\lambda_{ab}= 2756$ | $\kappa^{H//ab}=\gamma\kappa^{H//c}= 287.1$ |

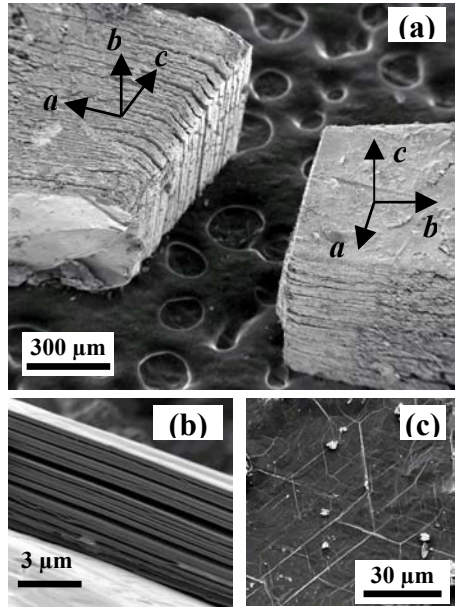

FIG. 1.

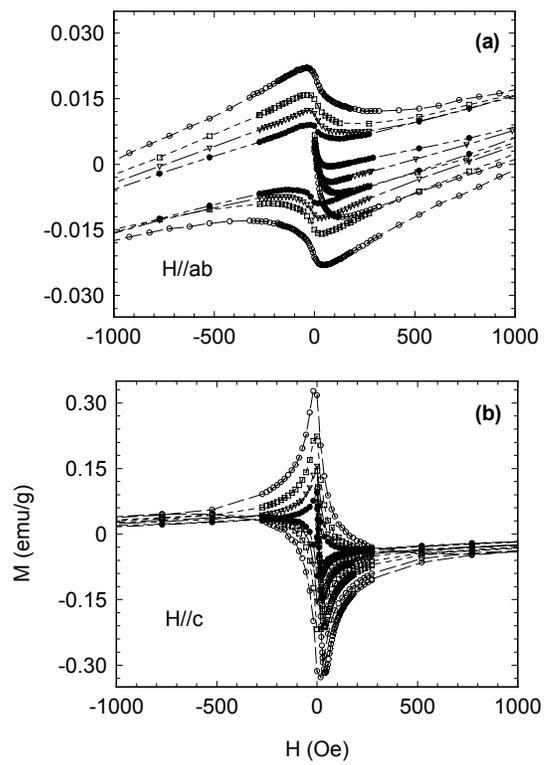

FIG. 2.

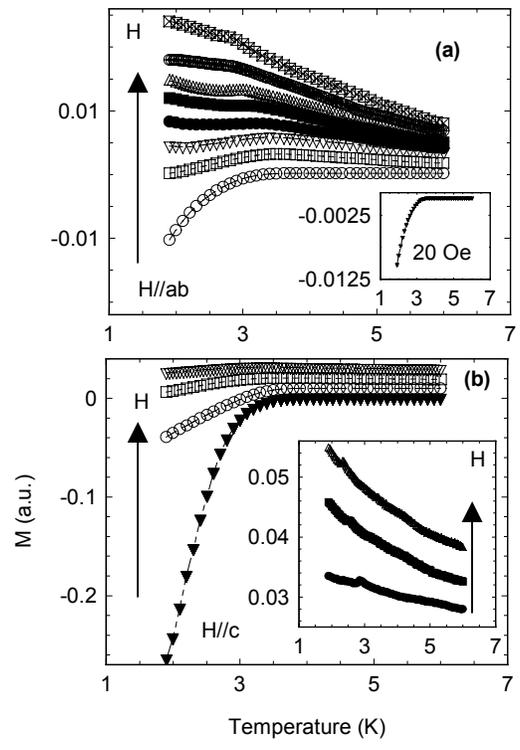

FIG. 3.

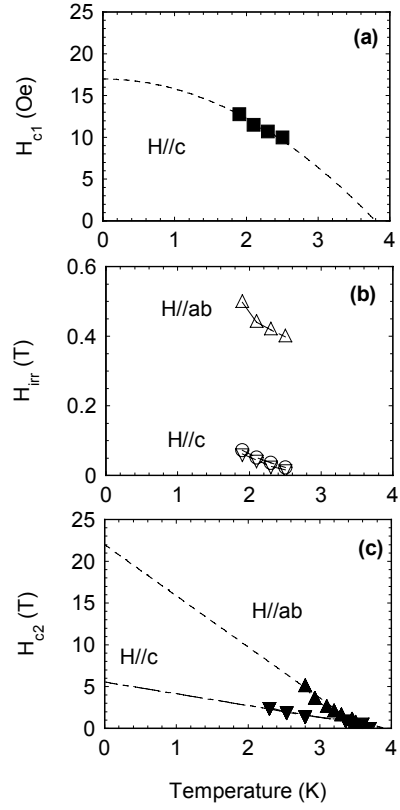

FIG. 4.